# Electronic transport properties of intermediately coupled superconductors: PdTe$_2$ and Cu$_{0.04}$PdTe$_2$

M. K. Hooda and C. S. Yadav*
School of Basic Sciences, Indian Institute of Technology Mandi, Mandi-175005 (H.P.) India
*Email: shekhar@iitmandi.ac.in

**Abstract**

We have investigated the electrical resistivity (*1.8 - 480 K*), Seebeck coefficient (*2.5 - 300 K*) and thermal conductivity (*2.5 - 300 K*) of PdTe$_2$ and *4%* Cu intercalated PdTe$_2$ compounds. Electrical resistivity for the compounds shows Bloch-Gruneisen type linear temperature (*T*) dependence for *100 K <T <480 K*, and Fermi liquid behavior ($\rho(T) \propto T^2$) for *T<50 K*. Seebeck coefficient data exhibit strong competition between Normal (*N*) and Umklapp (*U*) scattering processes at low *T*. The low *T*, thermal conductivity ($\kappa$) of the compounds is strongly dominated by the electronic contribution, and exhibits a rare linear *T* dependence below 10 K. However, high *T*, $\kappa(T)$ shows usual *1/T* dependence, dominated by *U* –scattering process. The electron–phonon coupling parameters, estimated from the low *T*, specific heat data and first principle electronic structure calculations suggest that PdTe$_2$ and Cu$_{0.04}$PdTe$_2$ are intermediately coupled superconductors.

PACS: 72.15.Eb, Electronic transport in metal

PACS: 74.25.F- Superconductivity

**Introduction**

The transition metal dichalcogenides (TMDCs) are extensively studied systems owing to their rich and diverse physical properties like superconductivity (SC), charge density wave (CDW), topological insulating states, and large magneto-resistance [1-7]. Among these TMDCs, interest in $PdTe_2$ has grown in past few years owing to its exciting physics and the advancement in experimental techniques [3,4,8-10]. The de Hass-van Alphen (dHvA) effect and magneto-resistance studies of $PdTe_2$, have shown that though the material exhibits multiband feature, the transport properties are dominated by the single band electronic structure with a very small contribution from other bands [10]. Recently, Fei *et al.* observed six conductive pockets in the dHvA measurements, and confirmed the non-trivial Berry phase originating from the hole pocket of the Dirac cone [8]. These results were supported by the dispersions in Angle Resolved Photoemission Spectroscopy (ARPES) studies [4,8,9]. The electronic structure calculation and ARPES studies by Y. Liu *et al.* showed the presence of Dirac cone surface states well below the Fermi level, instead of at The Fermi Level [4].

$PdTe_2$ has $CdI_2$ type hexagonal (P-3m1) crystal structure with the octahedral Pd-Te coordination, which is a notable exception to other superconducting TMDCs with trigonal prismatic coordination [11-14]. The *c/a* value of ~ *1.27* for $PdTe_2$ is much smaller in comparison to other $CdI_2$ type hexagonal TMDCs compounds [2]. The iso-structural compound $IrTe_2$ undergoes structural transition at 280 K and superconducts below 2.8 K upon 5 % Cu intercalation, after the suppression of structural anomaly [15]. Cu intercalation acts as electron dopant to $IrTe_2$ and enhances the interlayer orbital hybridization [16]. The Fermi liquid behavior has been observed in $IrTe_2$ up to 25 K [17]. Similarly, $PdTe_2$ was also reported to exhibit bulk SC below 1.7 K and intercalation of 5% Cu between the $PdTe_2$ layers enhanced the transition temperature to 2.6 K [3]. Although there are some reports available on the normal state electrical resistivity and Seebeck coefficient of $PdTe_2$, an elaborate account of the electronic transport (electric, thermal) properties of $PdTe_2$ and $Cu_{0.04}PdTe_2$ is lacking in literature [9,18-22]. Therefore we studied the electrical and thermal properties of the $PdTe_2$ and $Cu_{0.04}$ $PdTe_2$, to find out various normal state electronic parameters and transport scattering mechanism in these compounds.

In this report, we have discussed the nature of electric and thermal conduction mechanisms of PdTe$_2$ and Cu$_{0.04}$PdTe$_2$. Transfer of charge carriers upon Cu intercalation leads to enhancement in electronic thermal conductivity of PdTe$_2$. First principle electronic structure calculations and low temperature (*T*) specific heat data suggest intermediately coupled nature of SC in these compounds. The values of electron-phonon coupling constant ($\lambda_{ph}$) parameter *0.59* and *0.64* suggest that PdTe$_2$ and Cu$_{0.04}$PdTe$_2$ are intermediately coupled superconductors. The enhanced value of $\lambda_{ph}$ and density of states for Cu$_{0.04}$PdTe$_2$ leads to the enhanced value of superconducting transition temperature.

**Sample growth and Measurements:** Polycrystalline samples of PdTe$_2$ and Cu$_{0.04}$PdTe$_2$ were prepared by solid state reaction method by taking appropriate stoichiometric amount of required elements. The elemental mixture was first heated at *800 $^o$C* for *24* hours and subsequently treated at *500 $^o$C* for *7* days inside the evacuated sealed quartz tubes. The obtained compound was further grounded, pelletized and sintered at *500 $^o$C* for *24* hour. Room temperature X-ray diffraction (XRD) pattern (shown in figure 1) of both compounds confirm the hexagonal crystal structure (Space group: P-3m1). The lattice parameter '*a*' and '*c*' shows slight compression upon Cu intercalation (Table I). EDXS (Energy Dispersive X-ray Spectroscopy) measurements showed the compounds to be in stochiometric compositions of PdTe$_2$ and Cu$_{0.04}$PdTe$_2$.

Low *T*, electrical resistivity $\rho(T)$, Seebeck coefficient *S(T)*, thermal conductivity ($\kappa$), and specific heat (*C*) were measured using Quantum Design Physical Properties Measurement System (PPMS) and for high *T*, $\rho(T)$ from *300 K* to *480 K*, a homemade resistivity set-up was used. Magnetic measurements were performed using Quantum Design Magnetic Properties Measurement System (SQUID magnetometer).

**Results and Discussion**

**Electrical Resistivity**

The $\rho(T)$ measurements of the compounds are shown in figure 2. The residual resistivity ratios, RRR ($\rho_{300K}/\rho_{4K}$) in our polycrystalline compounds are *10* and *30* for PdTe$_2$ and Cu$_{0.04}$PdTe$_2$ respectively, which are comparable to that for the reported single crystals [3]. The room temperature (RT) resistivity of Cu$_{0.04}$PdTe$_2$ ($\rho_{300K} \approx 51.4$ $\mu\Omega$-cm) is slightly lower in comparison to pristine PdTe$_2$ (($\rho_{300K} \approx 60$ $\mu\Omega$-cm). It is to mention here that the RT resistivity value of PdTe$_2$ in our case is less than *70* $\mu\Omega$-cm reported for single crystal [3]. These $\rho_{(T=300K)}$ values are comparable to metallic PtTe$_2$, Ir$_{0.95}$Pt$_{0.05}$Te$_2$ and approximately two times lower than the other TMDCs like NbSe$_2$, NbS$_2$, TaS$_2$ and TaSe$_2$ [23-25]. The SC has been reported in these compounds at *1.69 K* and *2.6 K* for PdTe$_2$ and Cu$_{0.05}$PdTe$_2$ respectively [3,26]. Though we could not observe SC in PdTe$_2$ down to *1.8 K*, in our $\rho(T)$ measurement, Cu$_{0.04}$PdTe$_2$ shows SC below *2.4 K* shown in top inset of figure 2. The $\rho(T)$ follows $T^2$ dependence (Fermi liquid behavior) up to $T \sim 50$ *K* for both compounds (shown in the bottom inset of figure 2). Data was fitted using the expression $\rho(T) = \rho_0 + A_{e\text{-}e}T^2$. We obtained $A_{e\text{-}e}$ values of 2.55 nΩ-cm/K$^2$ and *2.64* nΩ-cm/K$^2$ and $\rho_0$ values of *5.32* μΩ-cm, and *1.45* μΩ-cm for PdTe$_2$ and Cu$_{0.04}$PdTe$_2$ respectively. These $A_{e\text{-}e}$ values are of same order as estimated by us from a single crystal PdTe$_2$ and Cu$_{0.05}$PdTe$_2$ [3]. Higher value of $A_{e\text{-}e}$ for Cu$_{0.04}$PdTe$_2$ indicates towards the stronger electron-electron scattering in the compound and higher value of effective mass $m^*$ in comparison with PdTe$_2$, The enhanced electron-electrons scattering can be understood in terms of the transfer of the charge carriers by intercalant copper to the PdTe$_2$ system. We have fitted resistivity data up to *480 K* using Bloch-Gruneisen (BG) formula

$$\rho_{BG} = 4R\left(\frac{T}{\theta_D}\right)^5 \int_0^{\theta_D/T} \frac{x^5 dx}{(e^x-1)(1-e^x)}$$ where $\theta_D$ is Debye temperature and $R$ is $T$ independent material parameter [27]. The values of $\theta_D$ obtained for PdTe$_2$ and Cu$_{0.04}$PdTe$_2$ are *116 K* and *108 K*. The values of the electron-phonon coupling coefficient ($A_{e\text{-}ph}$) obtained from the fit of $\rho(T) = \rho_0 + A_{e-ph}T$ for $100 < T < 480$ K are $1.92 \times 10^{-1}$ μΩ-cm/K and $1.65 \times 10^{-1}$ μΩ-cm/K for PdTe$_2$ and Cu$_{0.04}$PdTe$_2$ respectively. These values are larger in comparison to $\sim 10^{-3}$ μΩ-cm/K for good metals (Ag, Cu, Al etc.) [28]. The $\rho(T)$ saturates at high $T$ in $d$-band compounds when mean free path $l$, approaches the inter-atomic distance $d$, (Ioffe-Regel condition: $l \geq d$) [29]. The $\rho(T)$ values, *95* μΩ-cm for PdTe$_2$ and *80* μΩ-cm for Cu$_{0.04}$PdTe$_2$ at $T = 480$ K are within the Mooji limit of $\rho \approx 100 - 150$ μΩ-cm, above which materials show saturating behavior in $\rho(T)$ at higher $T$ [30].

**Seebeck coefficient**

The $S(T)$ of the compounds is shown in figure 3. Cu$_{0.04}$PdTe$_2$ shows negative value of $S$ in the range *2.5 - 300 K*, whereas for PdTe$_2$, $S(T)$ becomes positive above *140 K*. Cu intercalation between PdTe$_2$ layers enhances the chemical bonding across van der Waals gaps, and the hybridization between Te $p$-orbital and Cu $d$-orbitals, which lead to increase in carrier density [3]. The phonon drag minima which is observed at $T \sim 44K$ in PdTe$_2$, shifts to $\sim 35$ *K* in Cu$_{0.04}$PdTe$_2$. We have fitted low temperature, $S(T)$ data with the equation $S = A_{diff}T + B_{ph}T^3$, (figure 3b) and obtained electron diffusion coefficient ($A_{diff}$) and phonon drag coefficient ($B_{ph}$). The sign of $B_{ph}$ suggests that at low $T$, Umklapp ($U$) process dominates in Cu$_{0.04}$PdTe$_2$, whereas Normal ($N$) process dominates for PdTe$_2$. The values of $A_{diff}$ and $B_{ph}$ (shown in Table I) are comparable to Cu-Ag alloys[31]. In electron-phonon scattering, phonon wave vector $q$ may either bridges the filled region of $k$-space or empty region of $k$-space, leading to the positive or negative values of the phonon drag $S(T)$. Complex character of Fermi surface and various scattering processes give rise to negative phonon drag $S(T)$ in these compounds [9,21]. The shift in phonon drag minima for Cu$_{0.04}$PdTe$_2$ can be related to the increase in carrier concentration by copper intercalation and thus points towards the modification in the Fermi surface. Therefore, we have plotted $S/T$ versus $1/\rho$, known as Nordheim-Gorter (N-G) plots in figure 3c, 3d in the range $150<T<300\ K$ [32]. But contrary to our expectations, we did not observe any deviation from linearity in N-G plots, which suggests that there is no significant change in Fermi surface upon Cu intercalation. In a previous study, Liu Yan et al. have shown a very small change ($\Delta E \sim 16$ meV) in the binding energy upon the 5% Cu intercalation in PdTe$_2$ [9]. Since the N-G plot involves electrical resistivity, which depends on the various scattering processes, it is possible that variation from linearity is getting obscured by the noise in resistivity measurement. In high $T$ region (*150-300 K*), $S(T)$ varies linearly with $T$ for both compounds. Linear fit in this region gives the value of slope ($dS(T)/dT$) as *0.00298* μV/K$^2$ and *0.00288* μV/K$^2$ for PdTe$_2$ and Cu$_{0.04}$PdTe$_2$ respectively. The estimated values of Fermi energies ($E_F = \frac{-\pi^2 k_B^2}{6e}\left(\frac{dS}{dT}\right)^{-1}$) from the slopes, *4.11* eV and *4.25* eV for PdTe$_2$ and Cu$_{0.04}$PdTe$_2$ respectively, are comparable with the noble metals Cu (*7* eV), Ag (*5.49* eV) and Au (*5.53* eV) [28]. Using the values of Fermi energies observed from $S(T)$ data under the free electron gas approximation, estimated carrier concentrations are $3.78 \times 10^{22}$ electrons/cm$^3$ and $3.98 \times 10^{22}$ electrons/cm$^3$ for PdTe$_2$ and Cu$_{0.04}$PdTe$_2$, which are comparable to $\sim 1.8 \times 10^{22}$ electrons/cm$^3$ estimated from the Hall coefficient data of F. Fei *et al.*[8]. A slight increase in the carrier concentration for Cu$_{0.04}$PdTe$_2$, suggests that Cu intercalation donates electrons to parent PdTe$_2$ and increases carriers near the Fermi surface.

**Thermal Conductivity**

Figure 4 shows the $\kappa(T)$ of PdTe$_2$ and Cu$_{0.04}$PdTe$_2$ in the range $2 < T < 300\ K$. We observed broad maxima in temperature region 15 - 130 K, for both compounds. Such maxima occurs when $l_{ph-ph} \approx l_{ph-defects}$, where $l_{ph-ph}$ and $l_{ph-defects}$ are the mean free paths due to phonon-phonon and phonon-defects scatterings respectively. In this region wavelength of heat carrying phonons decreases and phonons are scattered by defects and impurities. Thermal conductivity of PdTe$_2$ is larger than Cu$_{0.04}$PdTe$_2$ in this region, which suggests that amounts of impurities and defects present in these two compounds are different and may be larger in PdTe$_2$.

We have observed a small dip in κ near T ~ 30 K for both compounds, which is quite intriguing and difficult to assign any explanation at the moment.

At low T, κ varies linearly with T below 10 K, instead of usual ~$T^3$ expected from the boundary scattering dominance. There are two possibilities for the occurrence of κ ∝ T behavior: (i) At low T, stacking faults and grain boundaries can give rise to Rayleigh type scattering from the core of the scattering center leading to linear T behavior (κ ∝ T) for the layered materials [33]. (ii) Under the free electron gas approximation, where low T scatterings are dominated by electrons and phonon wavelength ($\lambda_T$) is larger than electronic mean free path ($l_{elec}$), κ follows linear T dependence [33]. To sort out the above discussed possibilities about the linear behavior of κ below 10 K, we measured ρ(T) up to 100 K for both compounds on the pieces of same batch as used for κ(T) measurement. Using ρ(T) data, we estimated electronic thermal conductivity ($\kappa_e$) from Wiedemann-Franz law ($\kappa_e = LT/\rho$, where L is the Lorentz number). As shown in the inset (a) of figure 4, approximate 50 % of the total thermal conductivity ($\kappa_{tot} = \kappa_e + \kappa_L$) of PdTe$_2$ is accounted by the electronic contribution ($\kappa_e$). With Cu intercalation, lattice thermal conductivity ($\kappa_L$) decreases and $\kappa_e$ sharply rises to ~ 80 % of the $\kappa_{tot}$ value. Reduction in $\kappa_L$ occurs due to phonon scattering by smaller size Cu atoms and charge transfer in the van der Waals gap of PdTe$_2$. In low T limit, electrons are mainly scattered by impurities and defects as number of phonons are low, and $l_e$ becomes independent of T. Since at low T, electronic heat capacity varies linearly with T ($C_{el} \propto T$), the $\kappa_e = \frac{1}{3} C_v v_f l_e$ leads to $\kappa_e \propto T$, where $v_f$ is the Fermi velocity. Dominance of $\kappa_e$ contribution in $\kappa_{tot}$ leads to the linear T behavior of $\kappa_{tot}$ at low T.

At high T (>$\theta_D$), κ depends on mean free path of phonons; and electron-phonon scattering reduces the magnitude of $\kappa_e$, resulting in the 1/T dependence of κ (right inset of figure 4) for 100 K < T < 255 K. For T > 255 K, κ shows a weak T dependence where mean free path of phonons decreases on increasing temperature and becomes equal to the interatomic distance.

**Electronic Specific Heat**

Figure 5 shows the T dependence of specific heat (C) of the compounds. In the low T range, data is fitted with C/T = γ + β$T^2$ (shown in insets of figure 5). We obtained γ = 5.84 mJ/mol-K$^2$, β = 2.10 mJ/mol-K$^4$ for PdTe$_2$ and γ = 6.32 mJ/mol-K$^2$, β = 3.05 mJ/mol-K$^4$ for Cu$_{0.04}$PdTe$_2$. These γ values are comparable to

~7.02 mJ/mol-K$^2$ for iso-structural Pt$_{0.05}$IrTe$_2$ [34]. Using γ values, we estimated density of states $N(E_F) = 3\gamma/\pi^2 k_B^2$, to be ~ 2.47 states/eV-fu for PdTe$_2$ and 2.68 states/eV-fu for Cu$_{0.04}$PdTe$_2$. Enhanced value of $N(E_F)$ in Cu$_{0.04}$PdTe$_2$ indicates the carriers donation by Cu. The values of $\theta_D$ obtained from β values are ~140 K and ~ 136 K for PdTe$_2$ and Cu$_{0.04}$PdTe$_2$ respectively. These values are higher than the respective values of estimated 116 K and 108 K obtained from BG fit of the ρ(T) data in the range 4 K < T < 480 K.

**Electronic Band Structure Calculations**

We performed electronic structure calculations for PdTe$_2$ using Wien2k code within the generalized gradient approximation (GGA) for the electron correlations [35]. The basis set of size was chosen as R$_{MT}$K$_{max}$ = 7 and a 21×21×14 set of k-points was used for Brillouin zone sampling and density of states (DOS) plots. As seen from the figure 6, *Pd-4d* orbital electrons contribute more to the DOS at Fermi level than *Te-5p* orbitals. However, the total contribution from *Te* atoms dominates the overall contribution to DOS states at Fermi level. In a STM study on PdTe$_2$, the contribution of *Te* states was identified more than *Pd* contribution [22]. Bare density of states $N_{bs}(E_F)$ ~ 1.312 calculated from the first-principle study for PdTe$_2$ is ~ 1.88 times lower than the estimated experimental value. Renormalization factor λ ~ 0.88, calculated using $N(E_F)/N_{bs}(E_F) = 1 + \lambda$ is close to 0.89 for a PdTe$_2$-chain based compound, Ta$_4$Pd$_3$Te$_{16}$ [36]. The electron-phonon coupling constant ($\lambda_{ph}$) can be estimated from the McMillan formula [37] as;

$\lambda_{ph} = \frac{1.04 + \mu * \ln(\frac{\theta_D}{1.45 T_c})}{(1 - 0.62\mu)\ln(\theta_D/1.45 T_c) - 1.04}$; where μ (Coulomb repulsion parameter) is set as 0.13, $T_c$ used for PdTe$_2$ is 1.7 K and $\theta_D$ value is taken from heat capacity data. The estimated value of $\lambda_{ph}$ is 0.59, which points towards electron-nonphonon coupling strength parameter $\lambda_{nph} = \lambda - \lambda_{ph} = 0.29$. This value is half of the electron-phonon coupling strength $\lambda_{ph}$. Similarly $\lambda_{ph}$ for Cu$_{0.04}$PdTe$_2$ is estimated to be ~ 0.64. The $\lambda_{ph}$ values of 0.59 and 0.64 suggest that PdTe$_2$ and Cu$_{0.04}$PdTe$_2$ are intermediately coupled superconductors. The $\lambda_{ph}$ value depends on electronic DOS at Fermi level, phonon frequency, and electron-ion interaction [37]. The enhanced value of $\lambda_{ph}$ for Cu$_{0.04}$PdTe$_2$ is consistent with the increased DOS and lower value of $\theta_D$ (lower phonon frequency) and hence enhanced superconducting transition temperature.

**Conclusions**

We have shown electrical and thermal transport properties of PdTe$_2$ and 4 % Cu intercalated PdTe$_2$ compounds. Electrical resistivity of the compounds shows Fermi liquid behavior up to 50 K and follow Bloch-Gruneisen type linear behavior at high *T*. Seebeck coefficient and thermal conductivity show competition between Normal and Umklapp scattering processes at low *T*, and shows dominance of Umklapp process at high *T*. Cu intercalation enhances the contribution of $\kappa_e$ to total thermal conductivity ($\kappa_{tot}$) at low *T*. The enhanced electronic contribution below *10 K*, leads to linear *T* dependence. Low *T*, specific heat data and first principle electronic structure calculations suggest that $PdTe_2$ and $Cu_{0.04}PdTe_2$ are intermediately coupled superconductors.

**Acknowledgements**

We thank Uday Sood for help in sample preparation. We acknowledges Prof. A. K. Rastogi for the useful discussion and valuable comments. We acknowledge Advanced Material Research Center (AMRC), IIT Mandi for the experimental facilities. CSY acknowledges the IIT Mandi seed grant project IITMandi/SG/ASCY/29, and DST-SERB project YSS/2015/000814 for the financial support. MKH acknowledges IIT Mandi for the HTRA fellowship.

**References**

1. MOROSAN E., ZANDBERGEN H. W., DENNIS B. B. S., BOS J. W. G., ONOSE Y., KLIMCZUK T., RAMIREZ A. P., ONG N. P. and CAVA R. J., *Nat. Phys.*, **2** (2006) 544.
2. WILSON J. A. AND YOFFE A. D., *Adv. Phys.*, **18** (1969) 193.
3. RYU G., *J. Supercond. Nov. Magn.*, **28** (2015) 3275.
4. YAN L., ZHOU Z. J., LI Y., CHENG-TIAN L., AI-JI L., CHENG H., YING D., YU H., SHAO-LONG H., LIN Z., GUO-DONG L., XIAO-LI L., JUN Z., CHUANG-TIAN C., ZU-YAN X., HONG-MING W., XI D., ZHONG F. and ZING-XIANG Z., *Chin. Phys. Lett.*, **32** (2015) 6067303.
5. CHEN Y. L., ANALYTIS J. G., CHU J. H., LIU Z. K., MO S. K., QI X. L., ZHANG H. J., LU D. H., DAI X., FANG Z., ZHANG S. C., FISHER I. R., HUSSAIN Z. and SHEN Z. X., *Science*, **325** (2009) 178.
6. ALI M. N., XIONG J., FLYNN S., TAO J., GIBSON Q. D., SCHOOP L. M., LIANG T., HALDOLAARACHCHIGE N., HIRSCHBERGER M., ONG N. P. and CAVA R. J., *Nature*, **514** (2014) 205.
7. OOTSUKI D., PYON S., KUDO K., NOHARA M., HORIO M., YOSHIDA T., and FUJIMORI A., *J. Phys. Soc. Jpn.*, **82** (2013) 093704.
8. FEI F., BO X., WANG R., WU B., FU D., GAO M., ZHENG H., SONG F., WAN X., WANG B., WANG X. and WANG G., *Phys. Rev. B*, **96** (2017) 041201(R).
9. YAN L., JIAN-ZHOU Z., LI Y., CHENG-TIAN L., CHENG H., DE-FA L., YING-YING P., ZHUO-JIN X., JUN-FENG H., CHAO-YU C., YA F., HE-MIAN Y., XU L., LIN Z., SHAO-LONG H., GUO-

DONG L., XIAO-LI D., JUN Z., CHUANG-TIAN C., ZU-YAN X., HONG-MING W., XI D., ZHONG F. and XING-JIANG Z., *Chin. Phys. B*, **24** (2015) 067401.

10. Y. WANG, ZHANG J., ZHU W., ZOU Y., XI C., MA L., HAN T., YANG J., WANG J., XU J., ZHANG L., PI LI, ZHANG C. and ZHANG Y., *Scientific Reports*, **6** (2016) 31554.

11. FURUSETH S., SELTE K. and KJEKSHUS A., *Acta Chem. Scan.*, **19** (1965) 257.

12. THOMASSEN L., *Z. Physic. Chem. B*, **2** (1929) 349.

13. LYONS A., SCHLEICH D. and WOLD A., *Mat. Res. Bull.*, **11** (1976) 1155.

14. KIM W. S., CHAO G. Y. and CABRI L. J., *J. Less Comm. Met.*, **162** (1990) 61.

15. LI X., YAN J. Q., SINGH D. J., GOODENOUGH J. B. and ZHOU J. S., *Phys. Rev B*, **92** (2015) 155118.

16. KAMITANI M., BAHRAMY M. S., ARITA R., SEKI S., ARIMA T., TOKURA Y. and ISHIWATA S., *Phys. Rev. B*, **87** (2013) 180501(R).

17. CAO X., XIE W., ADAM PHELAN W., DITUSA J. F. and JIN R., *Phys. Rev. B*, **95** (2017) 035148.

18. RYAN W. G. and SHEILS W. L., *Phys. Rev. B*, **61** (2000) 12.

19. MYRON H. W., *Solid State Commun.*, **15** (1974) 395.

20. WESTRUM JR. E. F., and CARLSON H. G., *J. Chem. Phys.*, **35** (1961) 5.

21. KJEKSHUS A. and PEARSON W. B., *Can. J. Phys.*, **43** (1965) 438.

22. YU D. J., YANG F., MIAO L., HAN C. Q., YAO MENG-YU, ZHU F., SONG Y. R., ZHANG K. F., GE J. F., YAO X., ZOU Z. Q., LI Z. J., GAO B. F., LIU C., GUAN D. D., GAO C. L., QIAN D. and JIA J. F., *Phys. Rev. B*, **89** (2014) 100501(R).

23. MATSUMOTO N., TANIGUCHI K., ENDOH R., TAKANO H. and NAGATA S., *J. Low Temp. Phys.*, **117** (1999) 1129.

24. ZHOU S. Y., LI X. L., PAN B. Y., QIU X., PAN J., HONG X. C., ZHANG Z., FANG A. F., WANG N. L. and LI S. Y., *Euro. Phys. Lett.*, **104** (2013) 27010.

25. NAITO M. and TANAKA S., *J. Phys. Soc. Jpn.*, **51** (1982) 219.

26. RAUB CH. J., COMPTON V. B., GEBALLE T. H., MATTHIAS B. T., MAITA J. P. and HULL JR. G. W., *J. Phys. Chem. Solids*, **26** (1965) 2051.

27. BLATT F. J., *Physics of Electronic Conduction in Solids* (McGraw-Hill, NY1968). BASS, J., HELLWEGE, N. H. and OLSEN, J. J., *Electrical Resistivity, Thermoelectrical Power and Optical Properties* (Springer– Verlag 1st edition, Heidelberg, Berlin1985).

28. ASHCROFT N. W., AND MERMIN N. D., *Solid State Physics* (Harcourt College Publishers, Orlando, FL1976).

29. GUNNARSSON O., CALANDRA M., and HAN J. E., *Rev. Mod. Phys.*, **75** (2003) 1085.

30. MOOIJ J. H., *Phys. Status Solidi A*, **17** (1973) 521.

31. GUENAULT A. M., *J. Phys. F: Metal Phys.*, **4** (1974) 256.

32. NORDHEIM L. and GORTER C. J., *Physica*, **2** (1935) 383.

33. GREIG D., *Progress in Solid State Chemistry*, **1** (1964) 175.

34. FANG A. F., XU G., DONG T., ZHENG P. and WANG N. L., *Scientific Reports*, **3** (2013) 1153.

35. BLAHA P., SCHWARZ K., MADSEN G. K. H., KVASNICKA D. and LUITZ J.,WIEN2k, *An Augmented Plane Wave Plus Local Orbitals Program For Calculating Crystal Properties* (Vienna University of Technology, Austria (2001)

36. SINGH D. J., *Phys. Rev. B*, **90** (2014) 144501.

37. MCMILLAN W. L., *Phys. Rev.*, **167** (1968) 331.

**Figure Caption:**

Figure 1 (color online) XRD pattern of PdTe$_2$ and Cu$_{0.04}$PdTe$_2$ fitted with Rietveld refinement.

Figure 2 (color online). $\rho(T)$ versus $T$ data of PdTe$_2$ and Cu$_{0.04}$PdTe$_2$. Top left inset shows linear $T$ fitting of data. Bottom right inset shows $\rho(T) = \rho_0 + A_{e\text{-}e}T^2$ fitting up to *50 K* along with the superconducting transition.

Figure 3 (color online). (a) Seebeck coefficient $S(T)$ of PdTe$_2$ and Cu$_{0.04}$PdTe$_2$. (b) $S(T) = A_{diff}T + B_{ph}T^3$ fit to the low $T$ data. Nordheim-Gorter plot ($S/T$ versus $1/\rho$) for (c) PdTe$_2$ and (d) Cu$_{0.04}$PdTe$_2$.

Figure 4 (color online). Thermal conductivity ($\kappa$) versus $T$ data of PdTe$_2$ and Cu$_{0.04}$PdTe$_2$. Inset (a) shows electronic and total thermal conductivity below *100 K*. Inset (b) shows linear relation between $\kappa$ versus $1/T$ at high $T$.

Figure 5 (color online). Specific heat versus $T$ of PdTe$_2$ and Cu$_{0.04}$PdTe$_2$. Insets show $C_p/T$ vs. $T^2$ plot for PdTe$_2$ (bottom tight) and Cu$_{0.04}$PdTe$_2$ (top left).

Figure 6 (color online). Density of states for PdTe$_2$ showing total and partial density states for Pd and Te atoms. Dashed line represents the Fermi level.

Figure 1.

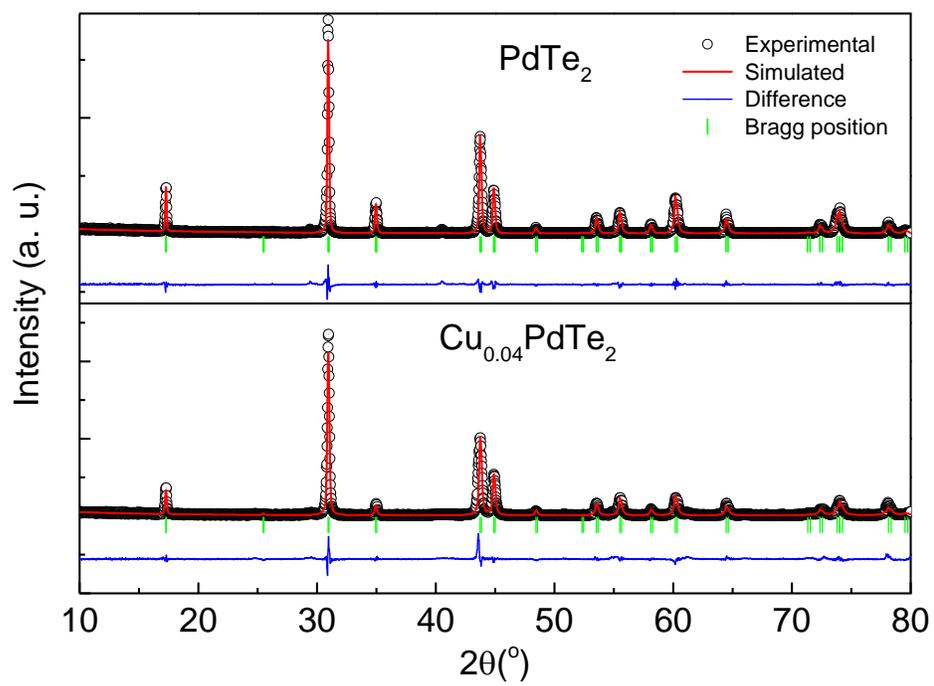

Figure 2.

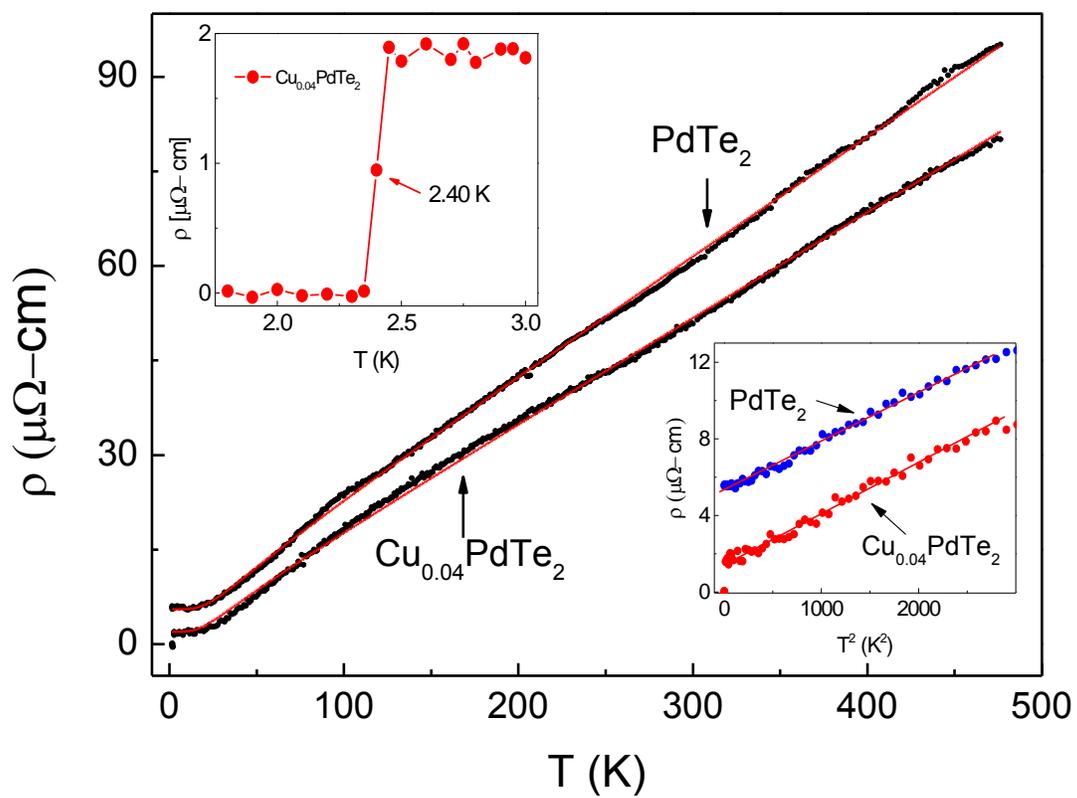

Figure 3.

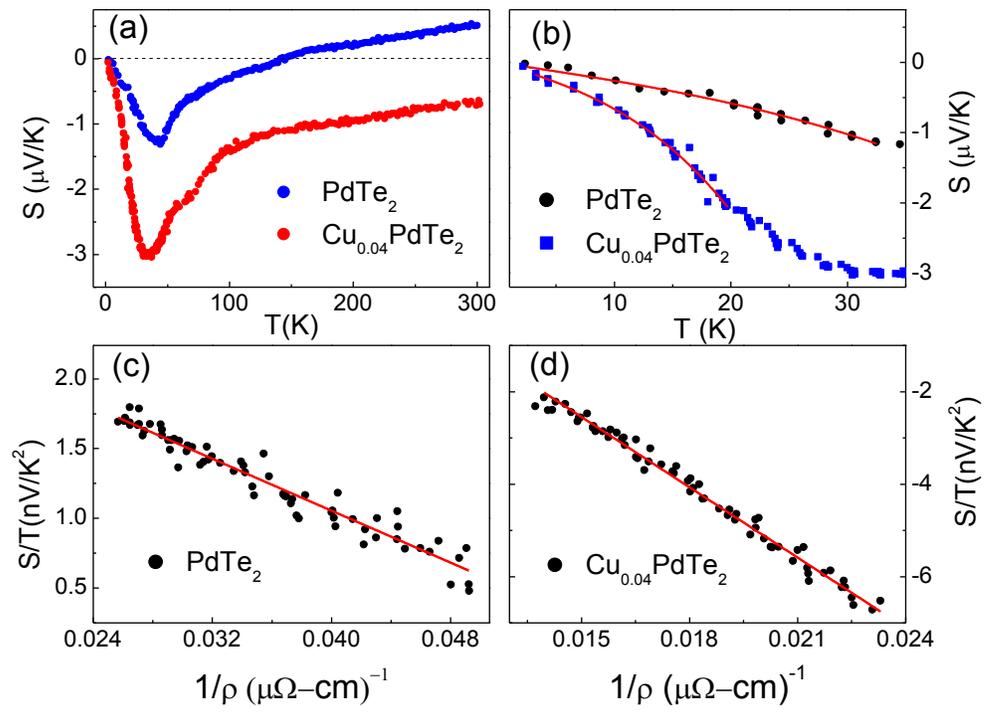

Figure 4.

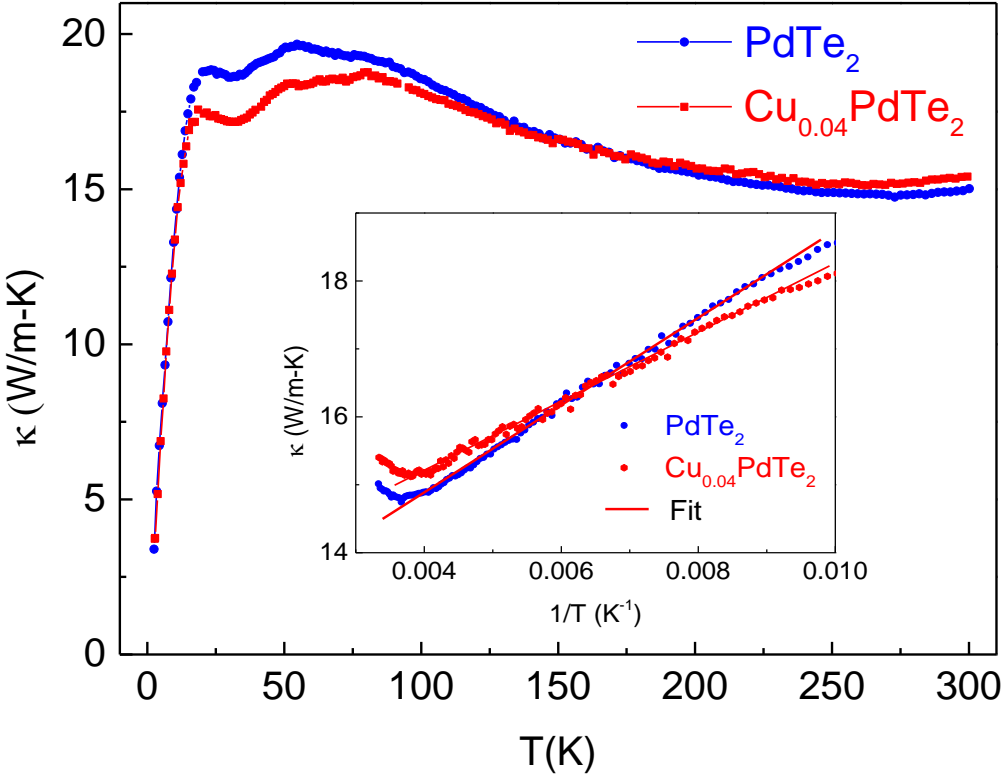

Figure 5.

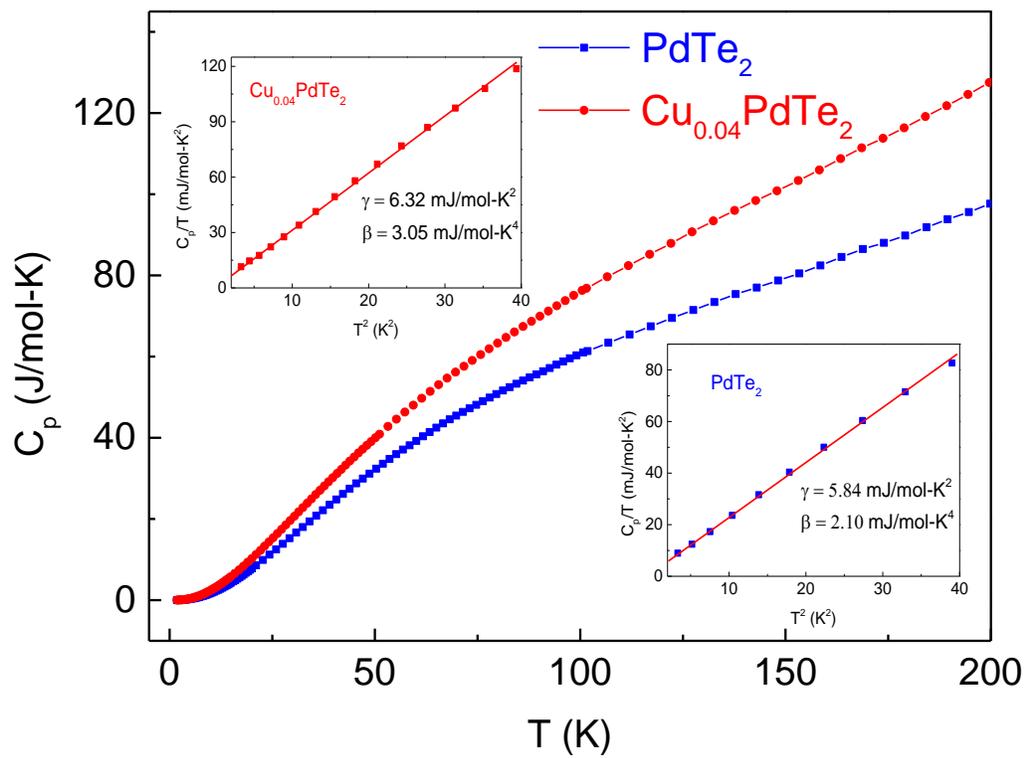

Figure 6.

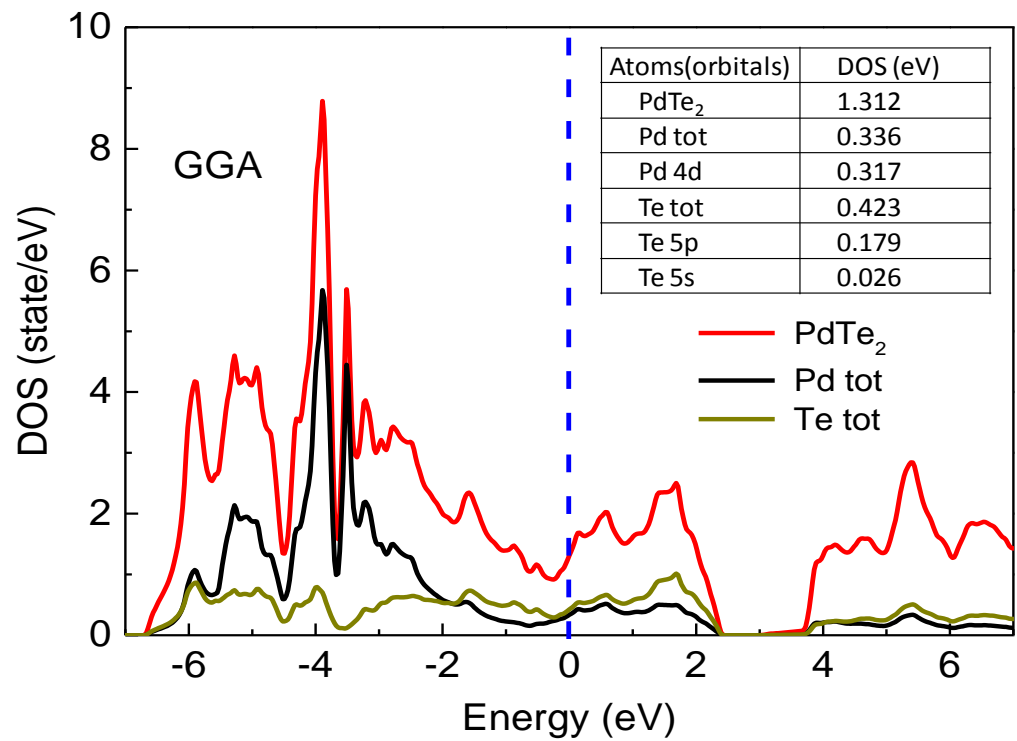

Table I

| Compound | a (Å) | c (Å) | V (Å$^3$) | $R_p$ | $R_{wp}$ | $\chi^2$ |
|---|---|---|---|---|---|---|
| PdTe$_2$ | 4.0368 | 5.1319 | 72.426 | 11.2 | 15.9 | 2.14 |
| Cu$_{0.04}$PdTe$_2$ | 4.0356 | 5.1315 | 72.378 | 9.26 | 14.4 | 3.20 |

Table II

| Compound | $A_{diff}$ (nV/K$^2$) | $B_{ph}$ (nV/K$^4$) |
|---|---|---|
| PdTe$_2$ | *-26.0* | *-0.009* |
| Cu$_{0.04}$PdTe$_2$ | *-53.1* | *0.132* |